\documentstyle[prb,twocolumn,aps,epsfig]{revtex}

\begin{document}
\title{Domain excitations in spin-Peierls systems.}
\author{Ariel Dobry}
\address{Departamento de F\'{\i}sica, Universidad Nacional de Rosario, \\
and Instituto de F\'{\i}sica Rosario, Avenida Pellegrini 250,\\
2000 Rosario, Argentina}
\author{David Ibaceta}
\address{Instituto de Astronom\'{\i}a y F\'{\i}sica del Espacio,\\
Casilla de Correos 67, Sucursal 28,\\
1428 Buenos Aires, Argentina.} 
\maketitle

\begin{abstract}
We study a model of a Spin-Peierls material consisting of a set of 
antiferromagnetic Heisenberg chains coupled with phonons and interacting
among them via an inter-chain elastic coupling. The excitation spectrum is
analyzed by bosonization techniques and the self-harmonic approximation. The
elementary excitation is the creation of a localized domain structure where
the dimerized order is the opposite to the one of the surroundings. It is a
triplet excitation whose  formation energy is smaller than the magnon gap. 
Magnetic internal excitations of the domain are possible and give the
further excitations of the system. We discuss these results in the context
of recent experimental measurements on the inorganic Spin-Peierls compound
CuGeO$_3$
\end{abstract}


The recent discovery of the first inorganic Spin-Peierls(SP) compound CuGeO$
_{3}$ has renewed the interest in the subject of quasi-one-dimensional
spin-phonon coupled systems\cite{Hase}. In spite of the intense experimental
and theoretical activity devoted to the study of this system neither the
excitations in the low temperature phase nor the mechanism of the
Spin-Peierls transition itself are well understood yet. Most studies have
used as a model Hamiltonian a Heisenberg antiferromagnetic chain with
alternating coupling and next nearest neighbors (nnn) interaction\cite
{Riedo,Castilla}. The Spin-Peierls transition is supposed to arise from a
competition between the elastic energy cost necessary to dimerize the chain
and the magnetic energy gain in this process. The dynamics of the phonons
are then supposed to be independent of the magnetic subsystem and an extreme
adiabatic approximation is assumed.

In spite of the success to reproduce both thermodynamical as well as
dynamical properties of Spin-Peierls systems by this approach, there are
now several indication that the Spin-lattice interaction in CuGeO$_{3}$ is
strong enough to make insufficient this approximation. No soft phononic mode
related to dimerization has been found\cite{Lorenzo}. This fact point toward
an order-disorder type transition where nonlinear excitations (domain wall)
different from phonons drive the structural transition. Otherwise, the phase
diagram of the system in presence of a magnetic field is now well
established. Above a critical value of the magnetic field the system
undergoes a transition from the uniform dimerized into an incommensurate
phase. The incommensurate lattice pattern has been measured by X-ray
experiments and interpreted as soliton lattice structure\cite{Kiry}.
The role of solitons in CuGeO$_{3}$ was indeed previously emphasized\cite
{Khomskii}. They could give an unified picture of the above features. They
could also be relevant to explain the rapid reduction of the SP temperature
by doping and the apparition of an antiferromagnetic phase\cite
{Khomskii,Fabrizio}.

It seems then natural to build the excitation spectrum of SP systems on the
basis of some kind of solitonic excitation (in this work we use the term
soliton in the general sense of a finite energy localized configuration
which is simultaneously magnetic and structural excitations, we call kinks
the 1D solitons). The present work gives some insight in this direction.

Moreover, recent neutron scattering studies have given a detailed
information about magnetic excitations in CuGeO$_{3}$\cite{Ain}. They show a
'double gap' structure, in addition to a dispersive triplet excitation there
is another gap which separates the first peaks from the band edge of a
continuum. This continuum has been interpreted as due to kink excitations
\cite{Ain}. It was stated that no free kinks could exist due to the
inter-chain coupling\cite{Uhrig}. The inter-chain coupling were taken into
account in previous works as providing a linear confined potential between
the kinks\cite{Khomskii}. This potential arises from a mean field
approximation to the inter-chain coupling. Particularly in a recent work a
ladder of bound kink-antikink states was shown for this model\cite{Affleck1}.

In this paper we study the formation of solitonic structures in SP system
including the three dimensional character of the phonon field. We use
bosonization techniques to account for the low energy magnetic excitations. We
show that the excitation gap of the system corresponds to create a
domain-like localized structure in which the dimerization is in antiphase to
the one in the bulk material. No independent kinks are possible. Instead of
that the walls of the domain always find an equilibrium situation depending
on the value of the inter-chain interaction. Internal excited states of the
domain are the higher excited levels of the system. They could be connected
with the continuum seen in CuGeO$_{3}$.

Let us consider a system of Heisenberg antiferromagnetic chains immersed in
the phonon field of the materials. The spin-phonon coupling arises from the
modulation of the magnetic exchange by the lattice motion. Therefore we will
focus on the following spin-phonon Hamiltonian:

\begin{eqnarray}
H=H_{ph}+H_{mg}
\label{H}
\end{eqnarray}
\begin{eqnarray}
\frac{H_{ph}}{J} &=&\sum_{i,j}\frac{{P_{i}^{j}}^{2}}{2M}
+\frac{K_{\parallel }%
}{2}(u_{i+1}^{j}-u_{i}^{j})^{2}+\frac{K_{\perp }}{2}%
(u_{i}^{j+1}-u_{i}^{j})^{2}  \nonumber \\
\end{eqnarray}
\begin{eqnarray}
\frac{H_{mg}}{J}=\sum_{i,j}(1+(u_{i+1}^{j}-u_{i}^{j})){\bf S}_{i}^{j}\cdot 
{\bf S}_{i+1}^{j}
\end{eqnarray}
we denote by i the site in a j-th chain. ${\bf S}_{i}^{j}$ are spin-1/2
operators of the $i,j$ ion. $H_{ph}$ represents our simplified model for the
phonons. It contains a scalar coordinate $u_{i}^{j}$ and its conjugate
momentum $P_{i}^{j}$ which are supposed to be the relevant ion coordinates
for the dimerization process. Notice that, in CuGeO$_{3}$, these are mainly
related with the displacement of the O ions. $K_{\parallel }$ and $K_{\perp }
$ are the longitudinal and transversal elastic couplings. We have included
only the coupling in one direction, generalization to the other direction
is straightforward and produces only a redefinition of the quantity ${\bf B}%
(j)$ of Eq. (\ref{B}). This transversal coupling gives the three dimensional
character to the phonons field which has been neglected in previous studies
of solitons in SP system\cite{NF}. It is essential for our considerations. 
The model Hamiltonian given by (\ref{H}) does not include magnetic
interaction between the chains. We will discuss latter on the possible 
effect of this
interaction on the excitation spectrum. Moreover we have not introduced a
nnn interaction between the spins of a chain. Even though, it play an
important role in reproducing several experimental properties in CuGeO$_{3}$
we do not expect any qualitative change of the results presented below which
are based on the solitonic excitations and not strongly affected by this
frustrated interaction\cite{Dobry,Feiguin}.

A similar model of the one given by Eq. (\ref{H}) were recently studied by
 Essler, Tsvelik and Delfino \cite{Essler} using rigorous field
theoretical results. However, as the authors of ref.\onlinecite{Essler}
have integrated out the lattice coordinates  and  neglected the 
dynamic of the phonons 
they did not take into account the mixed excitation we will discuss
in the present paper. We remark that in CuGeO$_3$ the scale of magnetic
and phononic  excitation are of the same order of magnitude suggesting that
 the elimination of the phonons degree of freedom is not suitable. 
 
Owing to the one-dimensional character for the magnetic interaction in (\ref
{H}) we can resort to the bosonization techniques to take into account the low
energy excitations of the systems\cite{Affleck2}. The spin variables are 
approximately represented by the bosonic fields $\phi^j(x)$ and its conjugated
momentum $\Pi^j(x)$. In addition we retain only the  phonon modes producing
a smooth deviation of the dimerized pattern.  So, we make the replacement 
$(-1)^iu_i^j \rightarrow u^j(x)$.  The effective Hamiltonian
($H_{eff}=H_{ph}+H_{mg}$) is given by:

\begin{eqnarray}
H_{ph}=\sum_j \int dx \{ \frac{{P^j(x)}^2}{2M}+2K_{\parallel } u^j(x)^2
\nonumber\\ + 
\frac{K_{\perp}}2 (u^{j+1}(x)-u^j(x))^2\}  \label{Hcont}
\end{eqnarray}
\begin{eqnarray}
H_{bos}=\sum_j \int dx\{\frac1{2\pi }[\frac{v_s}\eta (\partial _x\phi^j(x)
)^2+v_s\eta (\pi \Pi^j(x) )^2]\nonumber\\
+u^j(x)\sin \phi^j(x)\}  \label{Hbos}
\end{eqnarray}
$v_s$ is the spin wave velocity and $\eta$ the exponent of the correlation
functions\cite{Affleck2,NF}. For the isotropic Heisenberg model with nn
interaction we have $v_s=\pi/2$ and $\eta=2$.

We analyze the excitations of $H_{eff}$ by using a self-consistent
harmonic approximation(SHA) which has been shown to be reliable for
Spin-Peierls systems \cite{NF}. It is nearly connected to the semi-classical
methods widely used in the study of nonlinear field theory \cite{Rajaraman}.
In the SHA the fields $\phi ^{j}$ are split into a classical variable $\phi
_{0}^{j}$ and its quantum fluctuation $\hat{\phi ^{j}}$. 
Terms up to quadratic order in $\hat{\phi^{j}}$ are retained
and the annulment of the first order term is required
. The resulting equations are:

\begin{eqnarray}
- \frac{2v_s}\eta (\partial _x ^2 \phi^j_0)+ u^j e^{-\langle \hat{\phi^j}%
^2\rangle /2} \cos\phi^j_0=0  \label{eqmov1}
\end{eqnarray}
\begin{eqnarray}
\sin\phi^j_0 e^{-\langle \hat{\phi}^2\rangle /2} +(2 K_{\perp} +
4K_{\parallel}) u^j +  \nonumber \\
K_{\perp}(u^{j+1}+u^{j})=0  \label{eqmov2}
\end{eqnarray}
$\langle \hat{\phi ^{j}}^{2}\rangle $ is the ground state expectation
value and 
the last equation is the classical equation for the lattice coordinates $u^j$
. Uniform solutions of (\ref{eqmov1}) and (\ref{eqmov2})  correspond to the
homogeneous dimerized configuration of the chains. They are given by: 
\begin{eqnarray}
\phi^j_0=\frac{\pi}2 (mod. 2\pi) \;\;,\;\; 
u^j(x)=u_0 =
\frac{1}{\pi (4 K_{\parallel})^{3/2}}  \label{u0}
\end{eqnarray}
the transverse coupling is not effective in this case and the state  is the
same as the one of a single chain problem.  Excitations over this state are
given by: 
\begin{eqnarray}
\omega(k)=v_s (k^2+k_0^2)^{1/2} 
\label{omega}
\end{eqnarray}
they represent a band of triplet excitations separated  from the ground
state by the gap $\Delta=v_s k_0=(\pi u_0)^{2/3}$ . These magnon-like
 excitations are usually characterized as the elementary excitations
 of a Spin-Peierls system.

However, there is a lower excitation energy of the system which arises when
inhomogeneous solutions of (\ref{eqmov1}) and (\ref{eqmov2}) are allowed. To
take it into account we invert eq. (\ref{eqmov2}) to obtain $u^{j}(x)$
and then replace it in eq (\ref{eqmov1}). The result is: 
\begin{eqnarray}
u^{j}(x)=u_{0}\sum_{j^{\prime }}{\bf B}(j^{\prime }-j)\sin \phi
_{0}^{j^{\prime }}(x)
\label{uj}
\end{eqnarray}
\begin{eqnarray}
\partial _{x}^{2}\phi _{0}^{j}+\frac{1}{\xi ^{2}}\sum_{j^{\prime }}{\bf B}%
(j^{\prime }-j)\cos \phi _{0}^{j}\sin \phi _{0}^{j^{\prime }}=0
\label{system}
\end{eqnarray}
with: 
\begin{eqnarray}
{\bf B}(j^{\prime }-j)=\int_{-\pi }^{\pi }\frac{dk}{(2\pi )}\frac{\cos
(k(j^{\prime }-j))}{1+\epsilon \sin ^{2}\frac{k}{2}}
\label{B}
\end{eqnarray}
$\epsilon =\frac{K_{\perp }}{K_{\parallel }}$ is the relative inter-chain
elastic couplings and $\xi =v_{s}/\Delta $ will become the characteristic
 width of the domain wall(see below).

Eq. (\ref{uj}) and (\ref{system}) are our basic system of differential
equations to search solitonic excitations of the system
(each equation is labeled by the index $j$ of the chain). They should be
solutions of the system under the requirement of total finite energy
respect to the uniform configuration. The inter-chain elastic
 coupling strongly limits
the form of these solutions. For example, free kinks do not fulfill this
condition because they create an infinite zone where the transverse
couplings are active. The fields $u^{j}(x)$ and $\phi ^{j}(x)$ should go
as $x\rightarrow \pm \infty $ to the values they have in the uniform
configuration (\ref{u0}). The simplest solution satisfying these conditions
can be constructed as follows. We freeze the value of $\phi _{0}^{j}$ in all
the chains but the 0-th to its value in the homogeneous SP state, i.e. $\phi
_{0}^{j}=\pi /2$ for $j\ne 0$. The system (\ref{system}) then reduce to a
simple equation over the 0-chain. 
It corresponds to the classical equation 
for an effective one-chain problem with renormalized parameters and an
additional term favoring the dimerization phase of the neighbors chains. We
solve the equation by direct integration imposing the previously discussed
boundary conditions. The result is:

\begin{eqnarray}
t(x)\equiv
\sin (\phi_0^0)= 1\;\;-\;\;2\;\; {\cosh}^2 [x_0/\xi] \;\; {\rm %
sech} [( x-x_0)/\xi]  \nonumber \\
{\rm sech} [( x+x_0 )/ \xi]  \label{tdom}
\end{eqnarray}
with 
\begin{eqnarray}
x_0=\frac12 \log\left[\frac{1+{\bf B}(0)+ 2 \sqrt{{\bf B}(0)}} {1-{\bf B}(0)}%
\right]  \nonumber \\
{\bf B}(0)=1/\sqrt{1+\epsilon}
\end{eqnarray}

\begin{figure}[htb]

\epsfig{file=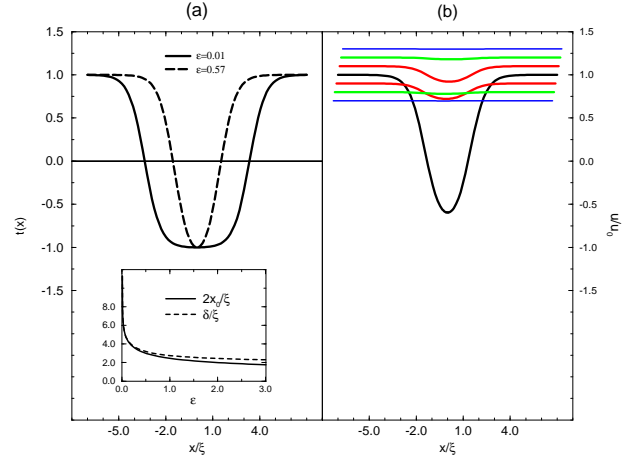,width=6cm,angle=-90}

\vskip .5 truecm

\caption{(a)function $t(x)$ defined in (\ref{tdom}).
 It gives the magnetic
profile of the domain excitation.
 The insertion shows the $\epsilon$%
-dependence of the parameter $2 x_0$ (related to the domain's size) and $%
\delta$ named the distance between the zeros of $t(x)$.
 b)Lattice pattern
corresponds to (\ref{udom}) for the $\epsilon=0.57$. We show the deformation
of the 0-chain and its neighbours chains.}

\label{fig1}

\end{figure}

This solution is shown in fig. (\ref{fig1}). We call it a domain. $2x_{0}$
measures the size of the domain configuration ( see the insertion of fig. (%
\ref{fig1})). It goes to infinity as the chain get decoupled ($\epsilon
\rightarrow 0$). In this limit, expression (\ref{tdom}) becomes a product of
two tanh-kink form. A small $\epsilon $ produces a rapid accommodation of
the walls one near the other in order to reduce the inter-chain energy. For
 a greater $\epsilon $, $2 x_{0}$ departs from the distance between the zeros
of $t(x)$ called $\delta$ in the inserted figure. As the walls could
not collapse this value never reaches the double of the width of the wall($%
2\xi $). Note that for each inter-chain coupling we have an equilibrium
distance. We do not find a linear confinement potential between the kinks as
it is assumed in other analysis\cite{Khomskii,Affleck1} of this problem.
 The domains turn out to be the elementary excitations of the
system. They will move as a whole when the translational invariance be
restore. This is a triplet excitation because its total $S_{z}$ could be 0
or $\pm 1$. Note that in the bosonic representation the total magnetization
is given by $S_{z}^{tot}=\frac{1}{2\pi }\sum_{j}(\phi ^{j}(\infty )-\phi
^{j}(-\infty ))$.

The lattice deformation pattern is obtained from (\ref{uj}): 
\begin{eqnarray}
u^{j}(x)=u_{0}{\bf B}(j)\sin \phi _{0}^{0}+(1-{\bf B}(j))u_{0}
\label{udom}
\end{eqnarray}
We show this configuration in fig. (\ref{fig1}b). The displacement amplitude
in the intermediate zone of the 0-th chain is smaller than $u_{0}$. The
distortion is not restricted to the 0-chain as for the bosonic field $\phi
_{0}$ . The neighbors chains are slightly distorted respect to the uniform
dimerized pattern. This distortion is indeed very weak as the result of the
rapid fall of the value of ${\bf B}(j)$ with j \cite{Dobry}. This behavior
together with the asymmetry seen between the magnetic and lattice pattern is
possible as a consequence of neglecting the inter-chain magnetic interaction.
We should expect that inclusion of this interaction will smear out the
deformation producing a kind of two (or three) dimensional domain with
dimerization pattern opposite to the one of the surrounding material. Note
that in our model spin-one excitations live in a chain but on turning off
the transversal magnetic exchange a triplet localized state could appear
exciting simultaneously several chains. Summarizing, a general solution of (%
\ref{system}) in presence of a magnetic interaction will be an enlarged
domain in the direction of the chains. 

\begin{figure}[htb]

\epsfig{file=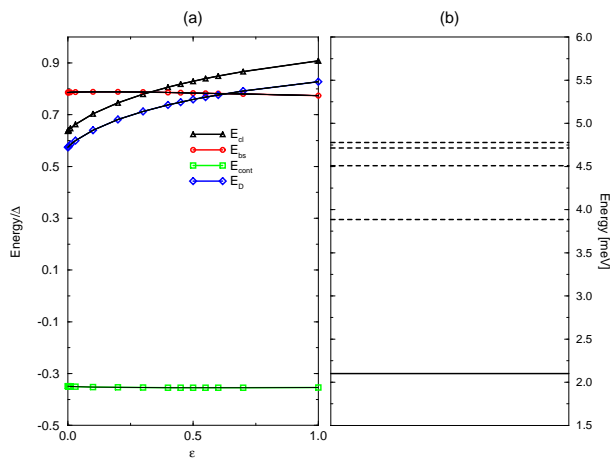,width=6cm,angle=-90}

\vskip .5 truecm

\caption{(a) Differents contribution to the domain energy ($E_d$). $E_{cl}$
classical energy, $E_{bs}$ contribution of the bound state of the potential (%
\ref{V0}), $E_{cont}$ contribution of the continuus(see text). (b) Energy
levels diagram for the parameter fitted for CuGeO$_3$. Dashed lines denote
the excited levels of the domain}

\label{fig2}

\end{figure}

We now turn to the study of the quantum states of the theory in terms of our
classical solution. The first contribution comes from the creation energy of
the domain. It is the difference of the values of the classical energy plus
the zero point energy of the fields obtained with and without the domain. The
classical energy is evaluated by replacing (\ref{tdom}) and (\ref{udom}) in (%
\ref{Hcont}) and (\ref{Hbos}) and subtracting the energy of the uniform
configuration. After a long but straightforward calculation we obtain: 
\begin{eqnarray}
E_{cl}=-\frac{1}{8\pi K_{\parallel }}[{\bf B}(0)I_{2}+2I_{1}]
\label{Ecl}
\end{eqnarray}
with: 
\begin{eqnarray}
I_{1} \equiv \frac{4}{\sqrt{{\bf B}(0)}%
}x_{0}  \nonumber \\
I_{2} \equiv -\frac{2}{\sqrt{{\bf B%
}(0)}}\left[ \frac{2}{\sqrt{{\bf B}(0)}}-2(1+\frac{1}{{\bf B(0)}}%
)x_{0}\right]
\nonumber 
\end{eqnarray}
The $\epsilon $-dependence of this classical energy in shown in fig. (\ref
{fig2}a). For $\epsilon \rightarrow 0$ we have $E_{cl}=\frac{1}{2\pi
K_{\parallel }}$ which is twice the classical creation energy of a kink in
the single chain problem (expressions (4.10) and (4.11) of ref.%
\onlinecite{NF} with the appropriate redefinitions of the constants). When 
the inter-chain coupling is switch on, the walls get closer and its total
energy slightly increases. Note that in fig.(\ref{fig2}a) we are showing the
energies in units of the magnon gap $\Delta $, therefore for any reasonable
choice of $\epsilon $ it will be more favorable to create a domain rather
than to a excite a magnon. This result is at the hearth of our claim that
the domains are the lower energy excitations of SP systems and will survive
when quantum fluctuation are included.

We now go to the inclusion of these quantum fluctuations. The fluctuation
operator is the quadratic form in the fields $\hat\phi^j$ of
 the SHA Hamiltonian.
Its eigenvalue problem corresponds to a set of Schroedinger-like equations 
for the following potentials:
\begin{eqnarray}
V^{0}(x)=({\bf B}(0)(t(x)-1)+1)t(x)
\label{V0}
\end{eqnarray}
\begin{eqnarray}
V^{j}(x) &=&{\bf B}(j)(t(x)-1)+1  \label{Vj} 
\end{eqnarray}
Once the eigenvalues ($ E_{\lambda }$) are known
 the frequency oscillations in presence of
the domain could be computed as $\omega _{d}(\lambda )=\Delta \sqrt{%
E_{\lambda }}$. The quantum correction to the classical energy (\ref{Ecl})
is the difference between the sum of the zero point energies of these
oscillators and the ones given by (\ref{omega}) i.e. in absence of the
domain (the cutoff dependence is eliminated when the term containing $%
\langle \hat{\phi ^{j}}^{2}\rangle $ in the SHA Hamiltonian is included \cite
{NF}).

Equation (\ref{V0}) defines a 
quantum mechanics problem of a particle in a double well potential, whereas,
Eq. (\ref{Vj}) the one of weak wells. We have solved numerically 
the eigenvalues problem. The bound states has been obtained by means of a nodes counting
algorithm. Numerical details of the method will be given elsewhere. For the
double well potential (\ref{V0}), we have found two bound states
 (called $E_{1}$ and $E_{2}$). For $\epsilon \rightarrow 0$,
 the wells get far apart and the two levels collapse in a degenerate
 bound state with the same energy as the
bound state of the single kink problem\cite{NF}. The larger $\epsilon $, the
bigger is the splitting due to the tunneling between the wells.

Potentials (\ref{Vj}) have only one bound state near the border of the
continuum due to the weakness of the potential. We neglect its contribution
to the total creation energy as well as the slightly distortion of the
continuum. We will discuss later on the relevance of these bound states as
excited levels of the domain.

The contribution of the continuum levels of (\ref{V0}) could be computed
once the phase shift $\delta (k)$ of these states is known. We calculate its
by numerical integration of the equation starting from the origin and
matching its behavior to its asymptotic behavior at $x\rightarrow \infty $.
The final expression for the total contribution of these states is: 
\begin{eqnarray}
E_{cont} &=&\frac{\Delta }{2}\{\frac{1}{\pi }\int_{0}^{\infty }\frac{%
2-k\delta (k)}{\sqrt{k^{2}+1}}dk-1\}  \label{equ}
\end{eqnarray}
The last integral has been numerically evaluated. In fig. (\ref{fig2}a) we
show the $\epsilon $-evolution of the different pieces contributing to the
total creation energy($E_{d}$). The main dependence comes from the classical
energy. We find, again, that in the limit of decoupled chains our excitation
energy corresponds to the total energy of two independent kinks\cite{NF}.

What about CuGeO$_{3}$? As we have stated in the introduction,
 neutron scattering experiments show a clear dispersive excitation
 at 2.1 meV. This excitation was previously analyzed
 on the basis of a dimerized and frustrated Heisenberg
 chain\cite{Uhrig,Haas,Poilblanc}.
Taking into account the previous results we associate the energy gap with
 the energy formation of a domain.
 To fix parameters we use
the experimentally fitted values $J=120$K and $u_{0}=0.042$. With these
values we have $\Delta =2.68$ meV larger than the gap. We can therefore
choose $\epsilon =0.57$ giving $E_{d}$ equal to the measure gap. For this
$\epsilon$ the walls are near and the domain seems as a local depression
 of the SP order (see fig.\ref{fig1}). Inclusion of nnn interaction will change
  the parameters. The next energy levels of
the system are internal excitations of the domain. They correspond to the
bound states previously discussed. Fig. (\ref{fig2}b) shows a level diagram
for the low energies excitation of the system. It is natural to associate
the transitions to the excited levels with the continuum seen in CuGeO$_{3}$%
. We recall that switching on the inter-chain magnetic coupling will
smoother of the domain deformation in the direction perpendicular to the
chains. Therefore we expect that the internal excitations will form a
continuous band.  Note also that  a two domain continuum is possible starting
 near $2 E_d$.
To give precise predictions of the effect of the domain
formation on the neutron scattering spectra the detailed information about
the dynamics correlation functions will be needed. Note in addition that as
our domain is referred to a given state of dimerization degeneration between
the states at $q=0$ and $q=\pi $ is expected.

Let us assume a level excitations diagram for CuGeO$_{3}$ as the one of fig.
(\ref{fig2}b). As a finite density of domain will be thermally created our
model predict transitions corresponding to rise the domain to its excited
levels. The first one should appear as a low energy peak at about $\omega
_{d}(1)=1.78$ meV in the optical response of the system and to increase its
intensity with the temperature up to $T_{SP}$.
 Recent inelastic light
scattering experiences\cite{Els} have identify a new resonance
in agreement with this prediction.
However, note that this transition were assigned in Ref 
\onlinecite{Els} to a three-magnon process on the externally dimerized
chain. More experimental and theoretical work will be need to decide between
these interpretations.

Finally, we have discussed in this paper only the magnetic excitations of
the system. However the domain is a mixed state between the spins and
phonons. So we expect that the phonon response will be sensitive to the
domain formations. The apparition of shoulder or satellite peaks
related to the frequency oscillations of the ions around the domain
should be expected\cite{greco}.

 In summary, we have identified a new excitation mode of Spin-Peierls
systems. It gives a natural extension of the usual singlet-triplet
excitation of the statically dimerized chain when the lattice relaxation is
allowed. As the excited spins are not confined to a dimer, internal
excitation are possible. We claim that these excitations could account for
some of the features seen in recent experiments.

We thank Th. Jolliceur, J. Riera, R. Migoni, C. Abecasis, O. Fojon,
E. Calzetta and J. Miraglia for useful discussions and A. Greco
 for illuminating discussions and critical reading of the manuscript.
 A. D. acknowledges   
Fundaci\'on Antorchas for financial support.


\begin{references}
\bibitem{Hase}  M. Hase, I. Terasaki and K. Uchinokura, Phys. Rev. Lett. 
{\bf 70}, 3651 (1993).

\bibitem{Riedo}  J. Riera and A. Dobry, Phys. Rev. B {\bf 51}, 16098 (1995).

\bibitem{Castilla}  G. Castilla, S. Chakravarty and V.J. Emery, Phys. Rev.
Lett. {\bf 75}, 1823 (1995).

\bibitem{Lorenzo}  J.E. Lorenzo {\it et al.}, Phys. Rev. B {\bf 50}, 1278
(1994).

\bibitem{Kiry}  V. Kiryukhin, B. Keimer, J. Hill, and A. Vigliante, Phys.
Rev. Lett. {\bf 76}, 4608 (1996); V. Kiryukhin {\it et al.}, Phys. Rev. B 
{\bf 54}, 7269 (1996).

\bibitem{Khomskii}  D. Khomskii et al., Czech. J. Phys. {\bf 46}, 3239
(1996); M. Mostovoy, D. Khomskii, Z. Phys. B {\bf 103}, 209 (1997)

\bibitem{Fabrizio}  M. Fabrizio and R. M\'{e}lin, Phys. Rev. B {\bf 56},
5996 (1997) 

\bibitem{Ain}  M. A\"{i}n {\it et al.}, Phys. Rev. Lett. {\bf 78}, 1560
(1997).

\bibitem{Uhrig}  G. S. Uhrig and H. J. Schulz, Phys. Rev. B {\bf 54}, R9624
(1996).

\bibitem{Affleck1}  I. Affleck, in {\em Dynamical Properties 
of Unconventional Magnetic Systems}
  (NATO ASI, Geilo, Norway, 1997) and  cond-mat/9705127.

\bibitem{NF}  T. Nakano and H. Fukuyama, J. Phys. Jpn {\bf 49}, 1679 (1980).

\bibitem{Rajaraman}  R. Rajaraman, Solitons and Instantons, Elsevier 1982.

\bibitem{Dobry}A. Dobry and J. Riera, Phys. Rev. B {\bf 55} R2912(1997).

\bibitem{Feiguin}  A. Feiguin, J. Riera, A. Dobry and H. A. Ceccatto,Phys.
Rev. B {\bf 53}, 14607 (1997).

\bibitem{Essler} F. Essler, A. Tsvelik, and G. Delfino
  Phys. Rev. B{\bf56}, 11001 (1997) 

\bibitem{Affleck2}  I. Affleck, Fields Strings and Critical Phenomena,
edited by E. Br\'{e}zin and J.Zinn-Justin (North-Holland, Amsterdam 1990),
pg. 563.

\bibitem{Haas} S. Haas and E. Dagotto, Phys. Rev. B {\bf 52}, R14396 (1995).

\bibitem{Poilblanc} D. Poilblanc, J. Riera, C.A. Hayward, C. Berthier and
M. Horvati\'c, Phys. Rev. B {\bf 18} R11941 (1997).

\bibitem{Els}  G. Els {\it et al}, cond-mat/9711047.

\bibitem{greco}  A.D. thanks A. Greco for point out this argument.

\end{references}
\end{document}